\documentclass[doublecol]{epl2}
\usepackage{amssymb}

\newcommand{\fig}[1]{Fig.~\ref{fig:#1}}
\newcommand{\eq}[1]{(\ref{eq:#1})} 

\usepackage{amsfonts}
\usepackage{amsmath}
\usepackage{graphicx}
\usepackage{subfigure}
\usepackage[hide]{todo}

\ifx \usenormalheadings \undefined
    
\else
    
\fi

\newcommand{\paperTitle}{Identifying influential spreaders and efficiently estimating infection numbers in epidemic models: a walk counting approach}
\newcommand{\shortpaperTitle}{Identifying influential spreaders and estimating infection numbers}
\newcommand{\theKeywords}{epidemic disease spreading, self-avoiding walks, SIR model, SIS model, complex networks, influential nodes}

\ifx\pdfoutput\undefined
\else
    \usepackage[bookmarks=false,colorlinks=false,citecolor=black,linkcolor=black,breaklinks=true,pdftex]{hyperref}
    \hypersetup{
        pdfauthor = {Frank Bauer, Joseph T. Lizier},
        pdftitle = {\paperTitle},
        pdfsubject = {Version 4.0; svn 197},
        pdfkeywords = {\theKeywords},
        pdfcreator = {LaTeX with hyperref package},
        pdfproducer = {pdftex},
        pdfstartview = {FitH}}
\fi

\begin{document}


\title{\paperTitle} 
\shorttitle{\shortpaperTitle} 
\author{Frank Bauer\inst{1}\thanks{E-mail: \email{bauer@mis.mpg.de}, \email{joseph.lizier@csiro.au}} \and Joseph T. Lizier\inst{1,2}}
\shortauthor{Frank Bauer \etal}
\institute{                    
  \inst{1} Max Planck Institute for Mathematics in the Sciences, Inselstrasse 22, D-04103 Leipzig, Germany\\
  \inst{2} CSIRO Information and Communications Technology Centre, PO Box 76, Epping, NSW 1710, Australia
}

\abstract{We introduce a new method to efficiently approximate the number of infections resulting from a given initially-infected node in a network of susceptible individuals. Our approach is based on counting the number of possible infection walks of various lengths to each other node in the network. We analytically study the properties of our method, in particular demonstrating different forms for SIS and SIR disease spreading (e.g. under the SIR model our method counts self-avoiding walks). In comparison to existing methods to infer the spreading efficiency of different nodes in the network (based on degree, k-shell decomposition analysis and different centrality measures), our method directly considers the spreading process and, as such, is unique in providing estimation of actual numbers of infections. Crucially, in simulating infections on various real-world networks with the SIR model, we show that our walks-based method improves the inference of effectiveness of nodes over a wide range of infection rates compared to existing methods.  We also analyse the trade-off between estimate accuracy and computational cost, showing that the better accuracy here can still be obtained at a comparable computational cost to other methods.}

\maketitle

\section{Introduction}
Epidemic spreading in
biological, social, and technological networks has recently
attracted much attention (see for instance
\cite{Cohen01,Albert00,Pastor-Satorras01, Dodds04, Chen2011, Newman02, Chung09}).
Most of these studies focus on the following question: Assume that
we first infect a randomly chosen individual of the network (\textit{patient zero}) - how likely
is it that a substantial part of the network will be infected?
In these earlier approaches the network was
considered as a whole and the role of patient zero
on the disease spreading process was neglected.

In this letter, we consider the role a single individual
plays in the spreading process rather than the global properties
of the network. It is of particular interest to identify the most
influential spreaders, and to do so without expensive simulations. This knowledge
could, for instance, be used to prioritise vaccinations for the most influential
spreaders. The number of neighbours of an individual is a
simple but crude approximation for an individual's influence,
and one has to take further
topological properties of the network into account to
understand the spreading process adequately
\cite{Kitsak10,klemm11a}. As such, \cite{Kitsak10,Chen2011,klemm11a} propose
different inference measures for a node's spreading influence such as
the $k$-shell decomposition, the local centrality measure or
eigenvector centrality.\footnote{For directed networks other methods exist for ranking the influence of nodes in different dynamical contexts (e.g. ranking
researchers according to influence on the scientific
community \cite{Lu11,
Zhou12, Radicchi09}.)}
All of these
approaches show strong correlations between their
measure of influence and the (simulated) number of infected nodes. There is
potential for improvement however in: i. considering network features encountered by longer infection walks, and ii. addressing the ultimate goal of
estimating the expected number of infections rather than merely obtaining
correlations.
Importantly, one can only predict whether an infection will be epidemic (i.e. a large portion of the network will be infected) or harmless from an estimate of infection numbers, not from correlation scores of an inference method alone.
In this letter, we present an
approach based on counting the number of potential
infection walks from a given initially infected individual. Our
method overcomes the above issues and allows us to consistently
estimate with very good accuracy the expected number of infections
from each patient zero.
Moreover, our method is very efficient and has low computational
costs.

\section{The general model}
We consider epidemic spreading on \textit{network} structures.
A complex network can be identified with a graph
$\Gamma=(V,E)$ \footnote{For simplicity, we
do not allow self-loops or
multiple edges.} (here $V$ is the vertex and $E$ is
the edge set) in an obvious way. We say that $i$ and $j$ are
neighbours, in symbols $i\sim j$, if they are connected by an edge.
In general, we deal with undirected graphs, though our formulae
are trivially extended for the directed case.
Often it is
convenient to describe a graph by its adjacency matrix $A=
(a_{ij})_{i,j = 1,\ldots, |V|}$ where the matrix element $a_{ij} = 1$
if $i$ and $j$ are neighbours and zero otherwise, and $|V|$ is the number of vertices. Furthermore, $d_i=\sum_ja_{ij}$ denotes the (out) degree of the vertex
$i$.

First we consider a generalization of the SIS
(susceptible/infected/susceptible)-model and the SIR
(susceptible/infected/removed)-model. In our model a disease is
spread in a network through contact between infected (ill)
individuals and susceptible (healthy) individuals. At a given time
step, each infected individual will infect each of its susceptible
neighbours with a given probability $0\leq \beta \leq 1$ (for
simplicity we assume that $\beta$ is the same for all pairs of
vertices - however generalization to variable $\beta_{ij}$ is straightforward).
An infected individual will be removed from
the network with probability $(1-\lambda)$ (modelling
either death or full recovery with immunity);
otherwise, an infected individual remains in the network
with probability $\lambda$ and remains susceptible to (re)infection
at the (very) next time step.
For $\lambda=0$ and
$\lambda=1$ this model reduces to the SIR and SIS model,
respectively.\footnote{\label{fn:discretisationDifference}
This is slightly different to the usual discrete-time SIS model, where infected individuals must return to susceptible at the next time step before reinfection is possible. In our interpretation, all non-removed nodes are susceptible (i.e. infected and susceptible are not mutually exclusive).
\todo{FB: One could also mention here that this comes from the discreteness of time - depends if you want to add this or not - see also my comment in the response letter.}
This difference allows us to mathematically generalise and study smooth transitions from the SIR to the SIS model, using the walk-counting approach.}

For a given network, we want to find the expected number of
infections given the person that was infected first. It is natural
to think of the spreading process in terms of infection
\textit{walks} in the corresponding graph.
The degree of a vertex is a first indicator of how many
individuals it will infect, however this
neglects all infection walks of length greater than one -
see also \cite{klemm11a} where the role of longer
infection walks in epidemic spreading is discussed and
numerical simulations were performed.\footnote{
This study used walk counts from a source node as a predictor of spreading efficiency. However, unlike the technique we present it did not convert those counts into a direct estimate of infection numbers, nor did it consider the appropriate types of walks (i.e. one must consider self-avoiding walks for the SIR model).}
Moreover such walks play a
very important role in the dynamics of complex networks.
In the following we will show that it is crucial to
also take longer infection walks into account in order to get
precise results.

The probability $p(i,j,k)$ that vertex $j$ is infected through a
walk of length $k$ given that the infection started at vertex $i$
can be written as
\begin{align}
p(i,j,k) = p_\mathrm{inf}(i,j,k) p_\mathrm{sus}(i,j,k),
\label{eq:p_i_j_k}
\end{align}
where $p_\mathrm{inf}(i,j,k)$ is the probability that vertex $j$ is
infected at time $k$ given that vertex $j$ is susceptible at time
$k$, and
$p_\mathrm{sus}(i,j,k)$ is the probability that vertex $j$ is
susceptible (i.e. not removed) at time $k$, both given that the infection started at vertex
$i$. (We refer to time here since the spreading
process is updated at discrete time steps; hence it
is equivalent to say that vertex $j$ is infected through a walk of
length $k$ or infected at time $k$).

For general
graph topologies it is difficult and expensive to calculate
the $p(i,j,k)$ exactly.
In the subsequent analysis we show how each of
$p_\mathrm{inf}(i,j,k)$ and $p_\mathrm{sus}(i,j,k)$ in turn can be approximated when
we make the following reasonable
simplification assumption: We assume that all infection walks
(of the same as well as different lengths) are
independent of each other, i.e. we treat them as if they have no
edges in common.\footnote{For walk lengths
less than or equal to two this assumption is not needed since all walks
are independent anyway.}
As our simulation results indicate, this is a
reasonable approximation. Using this \textit{independent walk}
assumption, we approximate:
$$p_\mathrm{inf}(i,j,k)\approx q_\mathrm{inf}(i,j,k)=1 -
\prod_{P_m\in\mathcal{P}(i,j,k)}(1-p(P_m)).$$ $\mathcal{P}(i,j,k)$
is the set of all walks from $i$ to $j$ of length $k$ and $p(P_m)$
is the probability that the infection takes place along the walk
$P_m$. This formula is easily obtained by noting that
$\prod_{P_m\in\mathcal{P}(i,j,k)}(1-p(P_m))$ is the probability
that $j$ is not infected at time $k$ given that it was susceptible
and that the infection started at vertex $i$. It is insightful to
rewrite the last equation in the following form:
\begin{equation}\label{pk}q_\mathrm{inf}(i,j,k)=1 -
\prod_{l=0}^{k-1}(1-\lambda^l\beta^k)^{s_{ij}^{k,l}}\end{equation}
where $s_{ij}^{k,l}$ is the number of walks from $i$ to $j$ of
length $k$ with $l$ repeated vertices, i.e. the number of walks
consisting of $k+1-l$ different vertices (including $i$ and $j$).

Let us have a closer look at the relationship between
$p_\mathrm{inf}(i,j,k)$ and $q_\mathrm{inf}(i,j,k)$. To properly
compute $p_\mathrm{inf}(i,j,k)$, one must compute infection
probabilities on each walk $P_m$ in some order, and properly
condition these infection probabilities on those of overlapping
previously considered walk in $\{ P_1, P_2, \ldots , P_{m-1}\}$.
This leads to properly conditioned infection probabilities $p(P_m
| P_1, P_2, \ldots , P_{m-1})$ and the expression:
$$p_\mathrm{inf}(i,j,k)=1 -
\prod_{P_m\in\mathcal{P}(i,j,k)}(1-p(P_m | P_1, P_2, \ldots ,
P_{m-1})).$$ Now, if infection has not already occurred on one of
these previously considered walks $\{ P_1, P_2, \ldots ,
P_{m-1}\}$, then $p_\mathrm{inf}(i,j,k)$ only differs from
$q_\mathrm{inf}(i,j,k)$ where $P_m$ has any overlapping edges with
$\{ P_1, P_2, \ldots , P_{m-1}\}$. Since infection did not occur
on any of these walks with shared edges, then some of the shared
edges for the walk $P_m$ may in fact already be closed (i.e.
dropping $p(P_m | P_1, P_2, \ldots ,
P_{m-1})$ below $p(P_m)$). This yields:
\begin{equation}
    \label{eq:plessq}
    p_\mathrm{inf}(i,j,k)\leq q_\mathrm{inf}(i,j,k) \text{ for all } k,
\end{equation}
i.e. the independent walk
assumption leads to an overestimation of $p_\mathrm{inf}(i,j,k)$.

Let us now study $p_\mathrm{sus}(i,j,k)$ in more
detail. We introduce $p_\mathrm{rem}(i,j,k) :=
1-p_\mathrm{sus}(i,j,k)$, (see footnote \ref{fn:discretisationDifference})
i.e. the probability that vertex $j$ is
removed at time $k$ given the infection started at vertex $i$.
For all $t\geq 1$, we have:
$$ p_\mathrm{rem}(i,j,t+1) =p_\mathrm{rem}(i,j,t)
        + p_\mathrm{sus}(i,j,t)p_\mathrm{inf}(i,j,t)(1-\lambda)$$
\begin{equation}
    \Longrightarrow p_\mathrm{sus}(i,j,t+1) -   p_\mathrm{sus}(i,j,t) = -(1-\lambda)p(i,j,t)
    \label{eq:psucUpdateRelation}
\end{equation}
Summing \eq{psucUpdateRelation} over all $t$ from $1$ to $k-1$ we obtain:
\begin{align}
    p_\mathrm{sus}(i,j,k) = 1 -
    (1-\lambda)\sum_{t=1}^{k-1}p(i,j,t),
    \label{eq:psucFromP}
\end{align}
where we used $p_\mathrm{sus}(i,j,1) = 1$. As before, we
use the independent walk assumption to approximate
$p_\mathrm{sus}(i,j,k)$ by $q_\mathrm{sus}(i,j,k)$, and also
$p(i,j,k)$ by $q(i,j,k)$ where we define:
\begin{equation}
    q(i,j,k):= q_\mathrm{inf}(i,j,k) q_\mathrm{sus}(i,j,k).
    \label{eq:q_i_j_k}
\end{equation}
So by analogy to \eq{psucFromP} we define:
\begin{equation}
    q_\mathrm{sus}(i,j,k)
        :=    1 - (1-\lambda)\sum_{t=1}^{k-1}q(i,j,t), \text{ \ \ } \forall k.
\end{equation}
We observe that $q_\mathrm{sus}$ satisfies an equation similar to
(\ref{eq:psucUpdateRelation}): $$q_\mathrm{sus}(i,j,k) =
q_\mathrm{sus}(i,j,k-1) - (1-\lambda)q(i,j,k-1).$$ We then
consider the connection between $p_\mathrm{sus}(i,j,k)$ and
$q_\mathrm{sus}(i,j,k)$. In order to investigate this we note that
walks of length one and two always satisfy the independence
assumption. Hence we have $p_\mathrm{inf}(i,j,k) =
q_\mathrm{inf}(i,j,k)$ and $p_\mathrm{sus}(i,j,k) =
q_\mathrm{sus}(i,j,k)$ for $k=1,2$.

Now we will prove by induction that $p_\mathrm{sus}(i,j,t) \geq
q_\mathrm{sus}(i,j,t)$ for all $t$. First we assume this is true
for $t = k$ (as demonstrated for $k=1,2$ above). Then considering
$t = k + 1$, combining \eq{p_i_j_k}  and  \eq{psucUpdateRelation}
we have:
\begin{eqnarray*}
p_\mathrm{sus}(i,j,k+1)
 &=& p_\mathrm{sus}(i,j,k)(1 - (1 - \lambda)p_\mathrm{inf}(i,j,k)) \nonumber \\
 &\geq & q_\mathrm{sus}(i,j,k)(1 - (1-\lambda)q_\mathrm{inf}(i,j,k)) 
\\&=&
q_\mathrm{sus}(i,j,k+1), 
\end{eqnarray*} where we used \eq{plessq} and our inductive assumption $p_\mathrm{sus}(i,j,k) \geq q_\mathrm{sus}(i,j,k)$.
Hence we conclude that:
\begin{equation}
    \label{eq:psusgreaterthanq}
    p_\mathrm{sus}(i,j,k) \geq q_\mathrm{sus}(i,j,k) \text{ \ \ } \forall k.
\end{equation}
That is, we systematically underestimate the probability
$p_\mathrm{sus}$ of being susceptible. Together with the
observation that we systematically overestimate the probability of
being infected in \eq{plessq}, these opposite
effects of our independent walks assumption may balance each other in \eq{q_i_j_k}.

We define the impact of vertex $i$ (the estimated number of
infections given that vertex $i$ was infected first) as
$$ I_i := \lim_{L\to \infty}{I_i(L)} = \lim_{L\to \infty}\sum_{k=1}^L\sum_jq(i,j,k).$$
$I_i$ counts the \textit{total} number of infections, and so is not required to converge if $\lambda > 0$ since then some of the vertices might be infected several times.
Alternatively some other studies define outbreak size as the number of nodes infected at least once, though this does not inform one as to whether the infection will die out or not.
(Note that if the nodes cannot be infected several times, i.e. for the SIR model,
both aforementioned quantities coincide.)
It is easy to verify that in considering only walks of length $1$, the impact of vertex $i$ is $I_i(L=1)=\beta d_i$. This shows that
the degree $d_i$ of vertex $i$ is the first
order approximation of the impact $I_i$.
In order to calculate the $q(i,j,k)$ we need to know all the
$s_{ij}^{k,l}$.
%
The calculation of $s_{ij}^{k,l}$ from $i$ to all $j$ can be completed in $O(D^k)$ steps (average case of counting walks along homogeneous out-degree $D$ nodes with independent edges), and is the computational-time bottleneck for our method. (We propose later an asymptotically more efficient calculation for the SIS case). Crucially, while this asymptotic scaling is the same as for simulating the disease spreading process, the constant factor of proportionality for our technique is smaller by several magnitudes.\footnote{
The expected number of evaluations $e$ per simulation consists of $D$ evaluations of disease spread to each neighbour plus the same expected number of evaluations $e$ per $D\beta$ infected neighbour (on average, in the sub-critical regime); i.e. $e=D+D\beta e$. One can solve $e = D / (1-D\beta)$, but it is more useful to write this to limited walk length $k$ as $O(D^k\beta^{k-1})$. Crucially, we require the number of repeat simulations to be $\gg 1/\beta^{k-1}$ for proper sampling, and new simulations are required for each $\beta$. These two requirements push the constant factor orders of magnitude beyond that for our technique since we only need to calculate the SAWs once for all $\beta$.}

In the following, we restrict ourselves to the special cases
$\lambda= 0,1$ where we obtain the SIR and SIS models.

\section{The SIS-model}
The SIS model corresponds to the case $\lambda=1$, i.e. where infected nodes
always become susceptible again (i.e. do not die or become immune).
Examples of SIS-type disease spreading include computer viruses
and pests in agriculture where the individuals (computer/crops) do
not develop immunity against the disease and hence can be
re-infected again.

To compute $q_\mathrm{inf}(i,j,k>1)$, one has to count the number $s_{ij}^k$ of
different walks of length $k$ between $i$ and $j$, i.e. the number
of possible infection walks with \textit{any} number of repeated
vertices $l$. Crucially,
$s_{ij}^k$ is given by the $ij$-th entry of the $k$-th power $A^k$
of the adjacency matrix $A$,
which is computed in low-order polynomial time, making our method asymptotically much more efficient than simulations for the SIS special case. By equation (\ref{pk}) the
probability $p(i,j,k)$ that $j$ is infected by $i$ through a walk
of length $k$ is then approximated by (with $\lambda=1$):
\begin{equation}
    \label{p} q(i,j,k) =  1-(1-\beta^k)^{s^k_{ij}},
\end{equation} where we used $q_{\mathrm{sus}}(i,j,k) =1$
since $\lambda=1$ here.
We obtain:
\begin{align}\label{eq:ISIS}
I_i^{SIS} = & \lim_{L \to \infty} \sum_{k=1}^L \sum_{j \in V} 1 -
(1-\beta^k)^{s^k_{ij}}
\end{align}
(using $0^0=1$ by convention if
we allow $\beta =1$). Again we point out that this expression
might not converge (particularly if $\beta$ is too large) since
individuals can be infected several times.
Such divergence has a meaningful
interpretation: i.e. that
the infection will remain forever in the network and will not die
out. In \eqref{eq:ISIS} we take arbitrarily long
walks into account since the vertices cannot develop
immunity against the disease and so no upper
bound for the maximal length of an infection walk exists.

For other diseases it is more natural to assume that
the vertices can develop immunity after infection.

\section{The SIR-model}
The SIR model corresponds to the
case $\lambda=0$, i.e. where infected nodes never become
susceptible again after infection as the individuals develop
immunity or die. As far as infection spreading
is concerned, they are considered removed (i.e. they cannot
spread the virus, nor be reinfected).
Examples of SIR-type disease spreading include most diseases spread among humans.
Since a vertex cannot be infected twice,
we have to modify our previous considerations appropriately. Instead of general walks we now
have to consider \textit{self-avoiding walks} (SAWs) or paths.
Indeed, it has been previously suggested that an understanding of
SAWs would be useful in epidemiology \cite{eub05}, though this
was not properly investigated. It is well
known that, compared to counting walks, it is much more difficult
to count SAWs in a graph \cite{Madras93}.
However, instead of explicitly calculating the number of
SAWs one can obtain the number of SAWs recursively \cite{hayes98}.
In particular, the
number of SAWs from $i$ to $j$ of length $k+1$ is given by:
$$s_{ij}^{k+1,0}(\Gamma) = \sum_{g: g \sim j \text{ in } \Gamma} s_{ig}^{k,0}(\Gamma\setminus j),$$
for $k \geq 1$ where $s_{ig}^{k,0}(\Gamma\setminus j)$ is the
number of SAWs from $i$ to a neighbour $g$ of $j$ (in $\Gamma$) of
length $k$ in the graph that is obtained from $\Gamma$ by removing
the vertex $j$. The adjacency matrix of the graph $\Gamma\setminus
j$ is obtained from the adjacency matrix of $\Gamma$ by deleting
the $j$-th row and column.

Noting that there cannot exist a SAW of length $k>|V|-1$,
we obtain
for the overall expected number of infected vertices starting from
vertex $i$ (with $\lambda=0$):
\begin{equation}
\label{eq:SIR} I_i^{SIR} = \sum_{k=1}^{|V|-1} \sum_{j\in V} \left(1 -
(1 - \beta^k)^{s_{ij}^{k,0}}\right)\left(1 - \sum_{t=1}^{k-1}
q(i,j,t)\right).
\end{equation}
We write $I_i^{SIR}(L)$ to represent estimates with the sum over paths $k$ limited to maximum path length $L$.


\section{Simulation results}
%
We provide a brief application of our technique to simulations of SIR spreading phenomena using: a. the social network structure generated from email interactions between employees of a university \cite{gui03} (giant-component with 1133 nodes and 5451 undirected edges, diameter 8); b. the structure of the C. elegans neural network \cite{white86,watts98} (297 nodes and 2345 directed edges) to demonstrate a directed network, and c. the collaboration network of the arXiv cond-mat repository \cite{new04} (giant-component with 27519 nodes, 116181 undirected edges, diameter 16) to demonstrate a larger network.

For each network, we compute estimates $I_i^{SIR}(L)$ from \eq{SIR} for maximum (self-avoiding) walk lengths $L=1$ to 7 (max. of 5 for the cond-mat network), with variable infection rate $\beta$, for each patient zero $i$.
To investigate the accuracy of these estimates, we also compute numbers of infections for each patient zero $i$ and $\beta$ as averages $S_i^{SIR}$ over 1000 simulations (10000 for the cond-mat network).
Furthermore, to compare the accuracy of inferences of the \textit{relative} effectiveness of each node, we have also measured the \textit{k-shell} \cite{Kitsak10} and \textit{eigenvector centrality} for each node in the network
(these measures were suggested as useful inferrers of relative spreading efficiency from each node in \cite{Kitsak10} and \cite{klemm11a}).
Using Java code on a
2.0 GHz Intel Xeon CPU E5-2650, the 10000 simulations $S_i^{SIR}$ for all nodes and $\beta$ values were completed
for the cond-mat network in around 2000 hours; our estimates $I_i^{SIR}(L)$ were completed for $L = 4$ and 5 in 30 and 60 minutes respectively;
while with Matlab scripts the degree and eigenvector centrality were completed in less than one second each and $k$-shell computed in less than 30 seconds.
We note that computation of the relevant walks $s_{ij}^k$ for SIS models is significantly more efficient than for SIR, since they can be directly computed from $A^k$ (as described earlier).
We chose to perform SIR simulations here in order to provide a greater computational challenge for our technique.

The extent to which our estimates accurately represent the \textit{relative} spreading effectiveness from each patient zero is examined via the correlation of estimates $I_i^{SIR}(L)$ to simulated results $S_i^{SIR}$ for the various networks in \fig{correlationsFigs}, as well as via their rank order correlations (defined in \cite{klemm11a}).
These figures demonstrate that our estimates $I_i^{SIR}(L)$ \textit{consistently} provide very accurate assessment of relative spreading effectiveness of the nodes over large ranges of $\beta$ \textit{for all networks examined}, in particular for $L>1$ and for $\beta$ values in the sub-critical, critical, and the lower-end of the super-critical spreading regimes.
(Critical spreading is defined as $\beta = \beta_c = 1 / \alpha_{max}$ \cite{klemm11a}, where $\alpha_{max}$ is the largest eigenvalue of the adjacency matrix $A$. \fig{correlationsFigs} indicates $\beta_c$ and also $\beta_{-3 dB}$ where 30 \% of the network is infected on average (upper super-critical regime) - the number of infections continue to increase very quickly beyond this $\beta$).

\newcommand{\corrFigWidth}{0.24\textwidth}
\begin{figure}[t]
        \subfigure[Email correlation]{\label{fig:emailCorrelation}\includegraphics[width=\corrFigWidth]{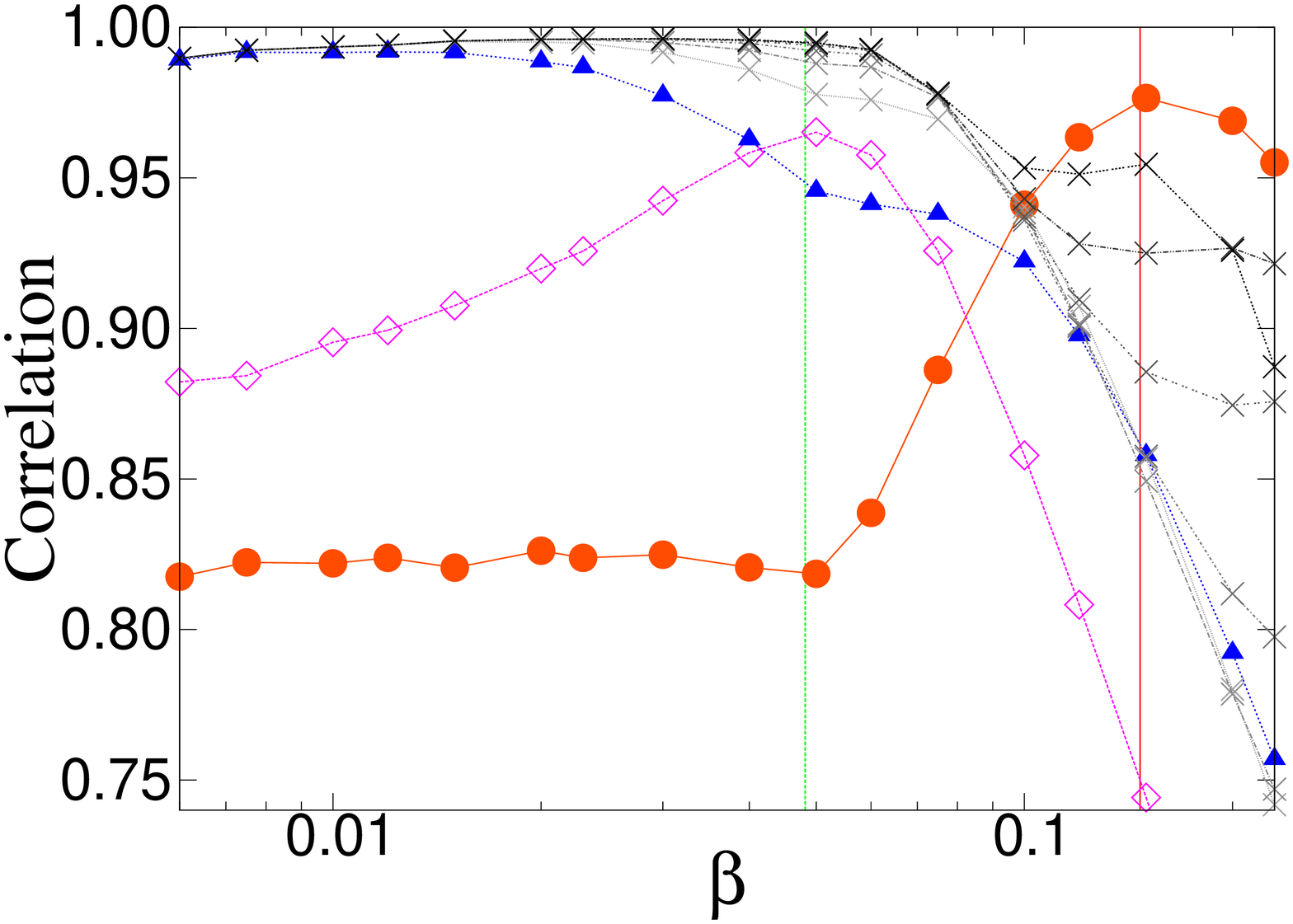}}
        \subfigure[Email rank order correlation]{\label{fig:emailRankOrder}\includegraphics[width=\corrFigWidth]{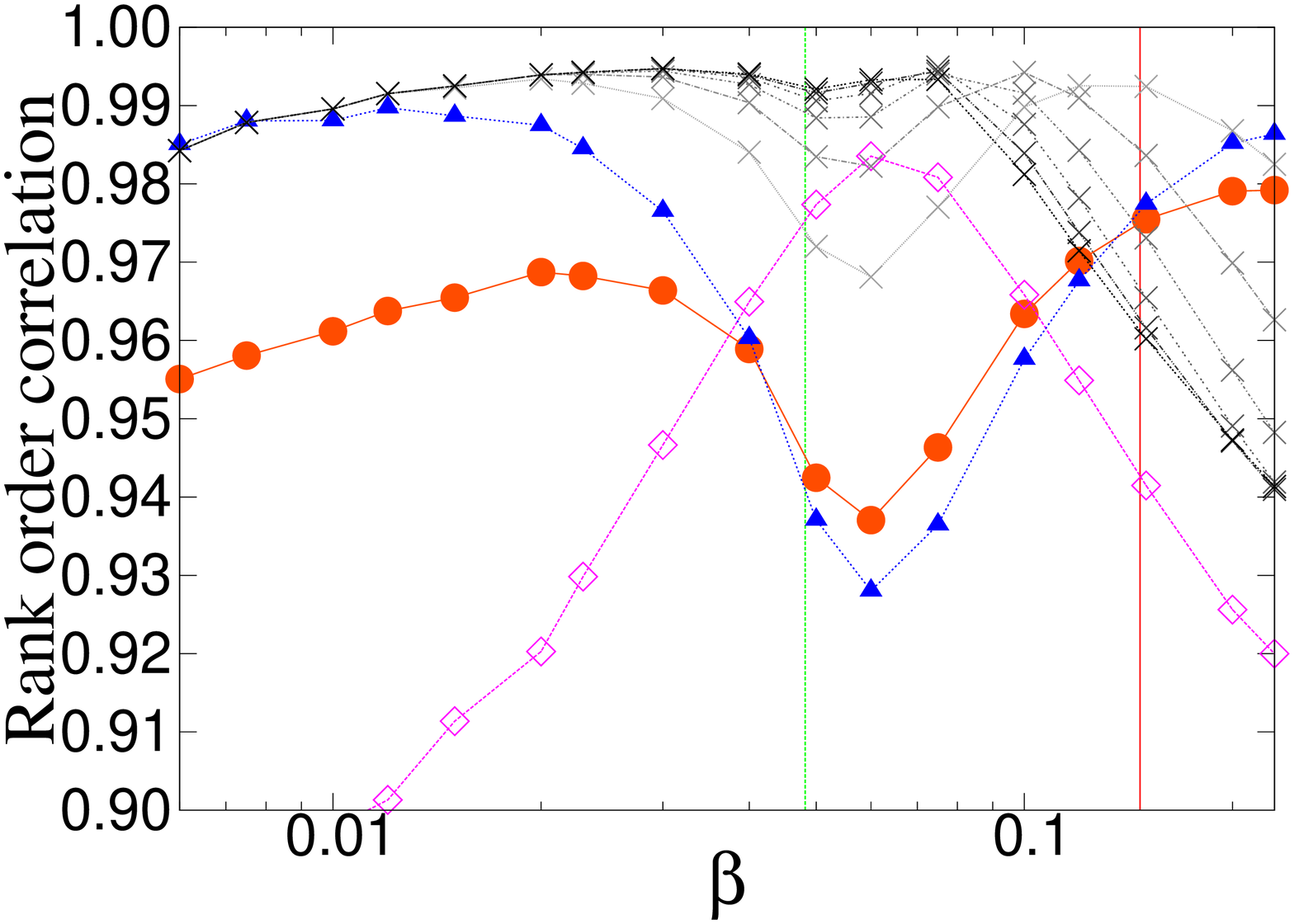}}
        \subfigure[C. elegans correlation]{\label{fig:celegansCorrelation}\includegraphics[width=\corrFigWidth]{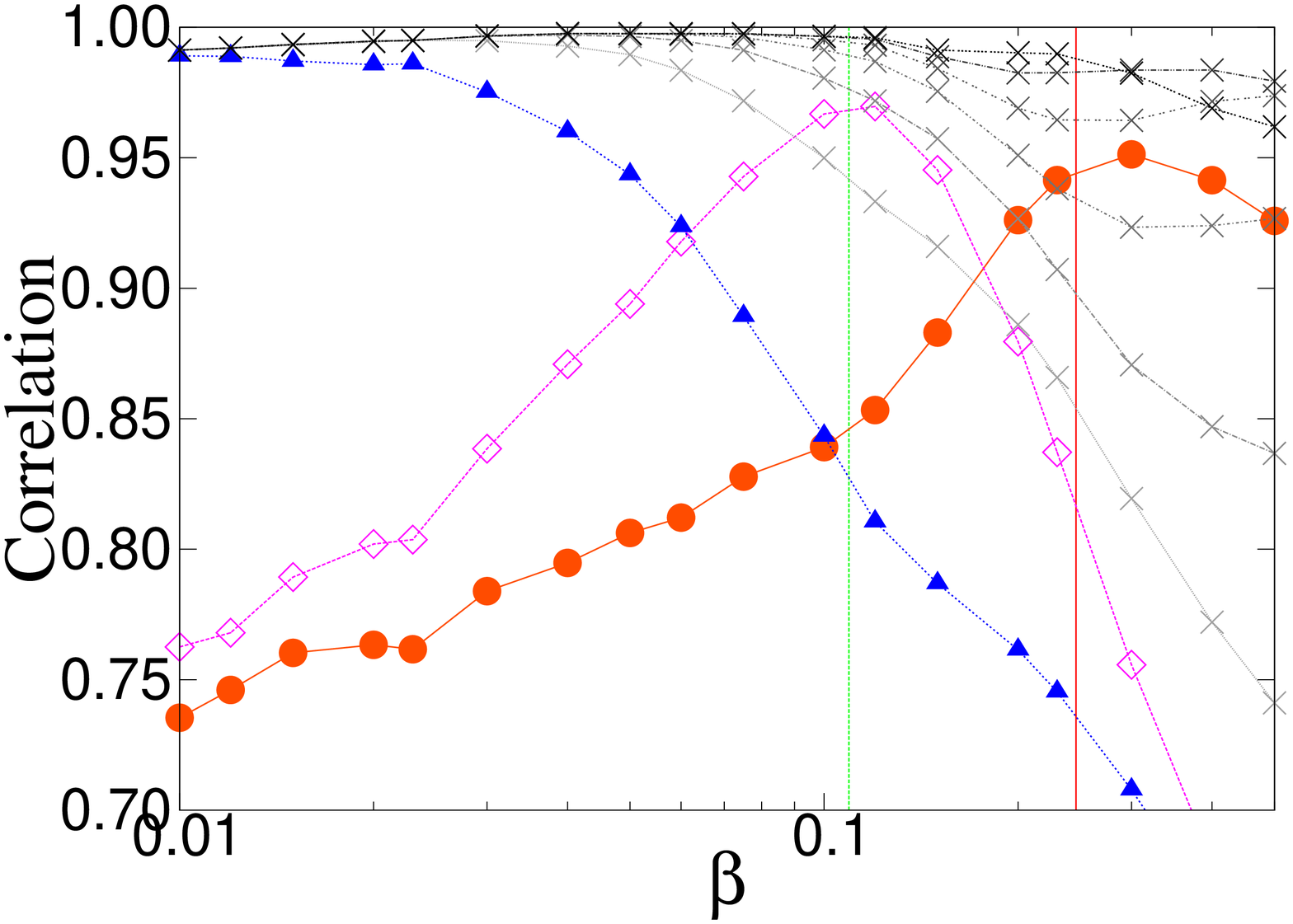}}
        \subfigure[C. elegans rank order corr.]{\label{fig:celegansRankOrder}\includegraphics[width=\corrFigWidth]{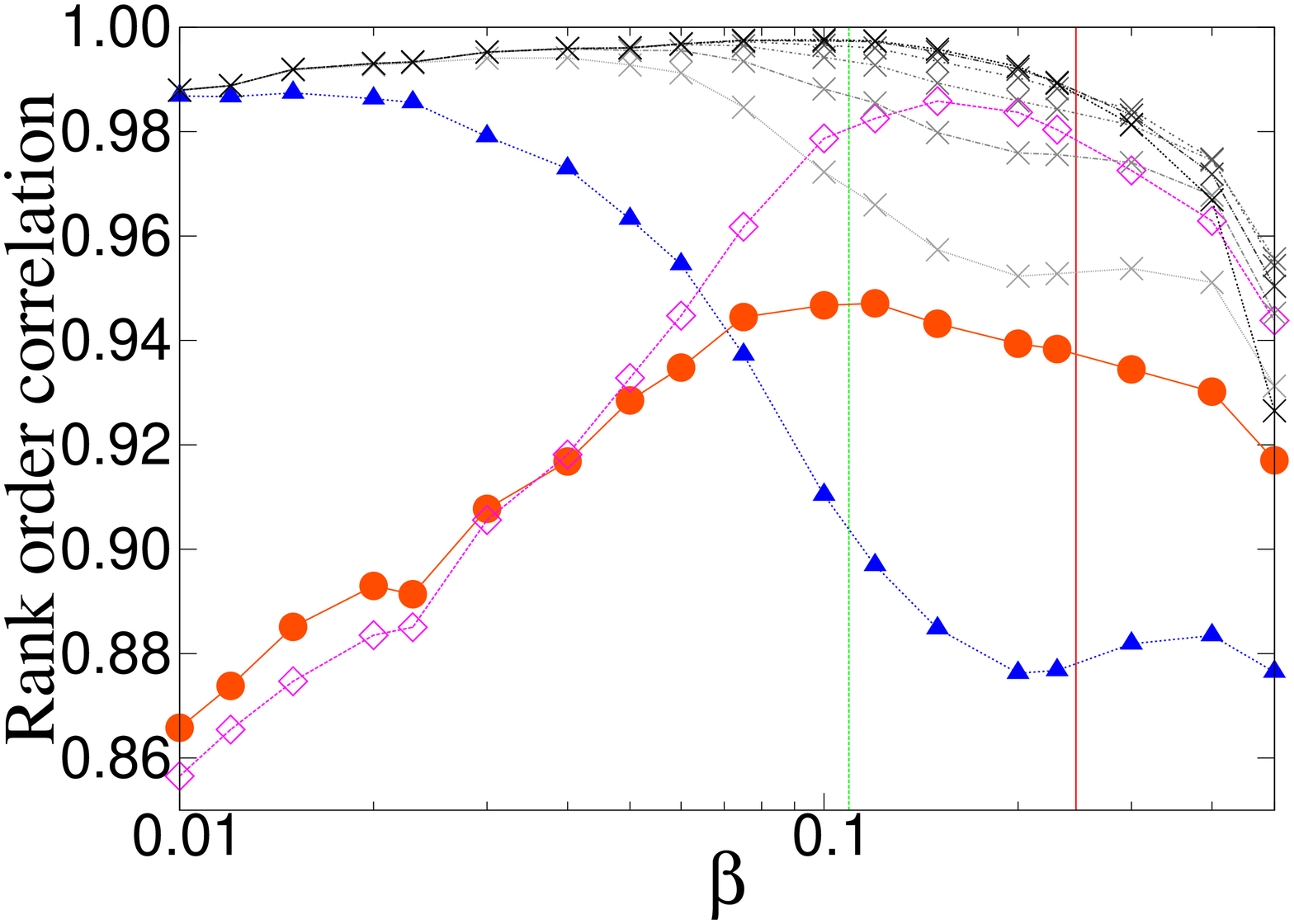}}
        \subfigure[Cond-mat correlation]{\label{fig:condmatCorrelation}\includegraphics[width=\corrFigWidth]{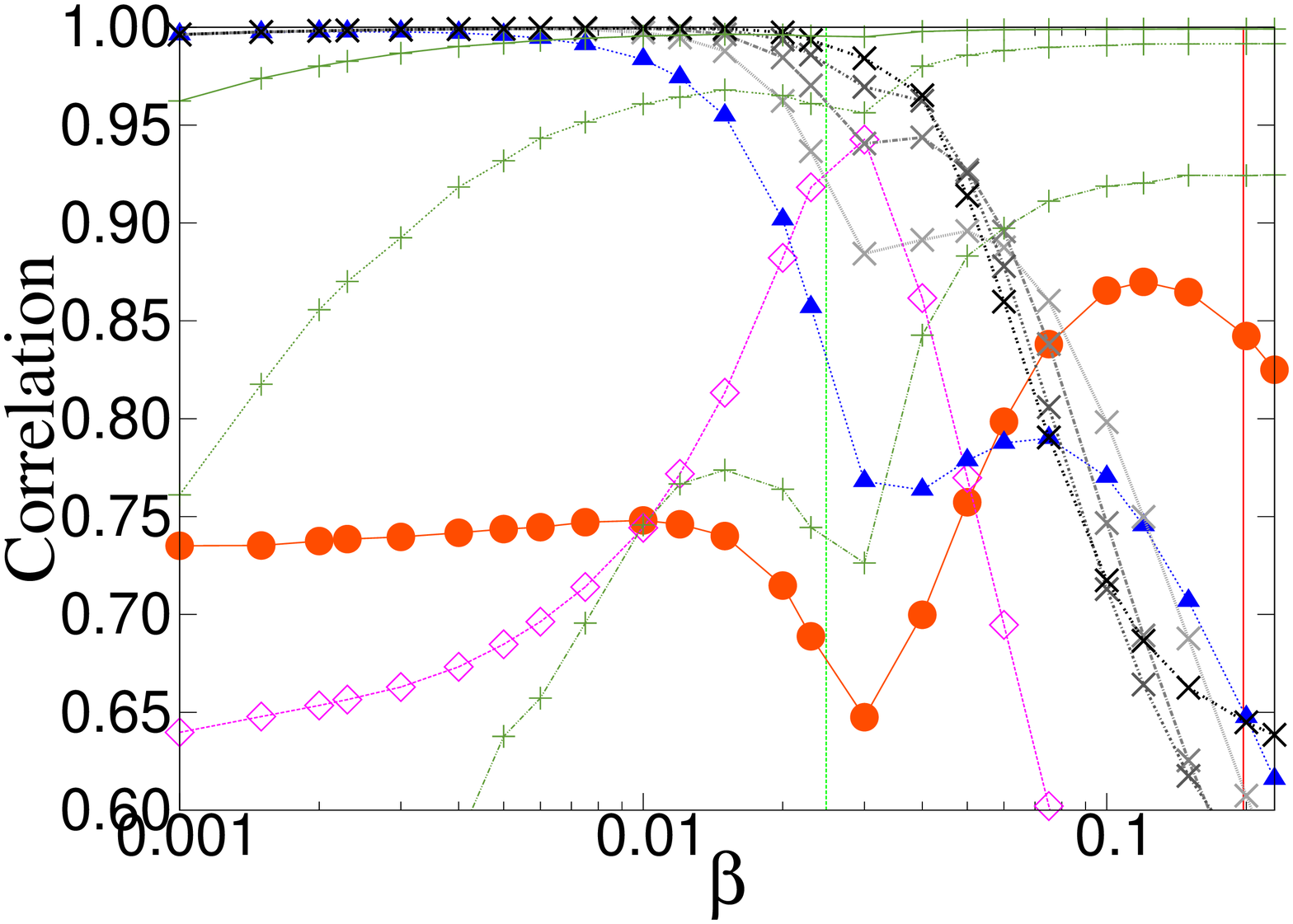}}
        \subfigure[Cond-mat rank order corr.]{\label{fig:condmatRankOrder}\includegraphics[width=\corrFigWidth]{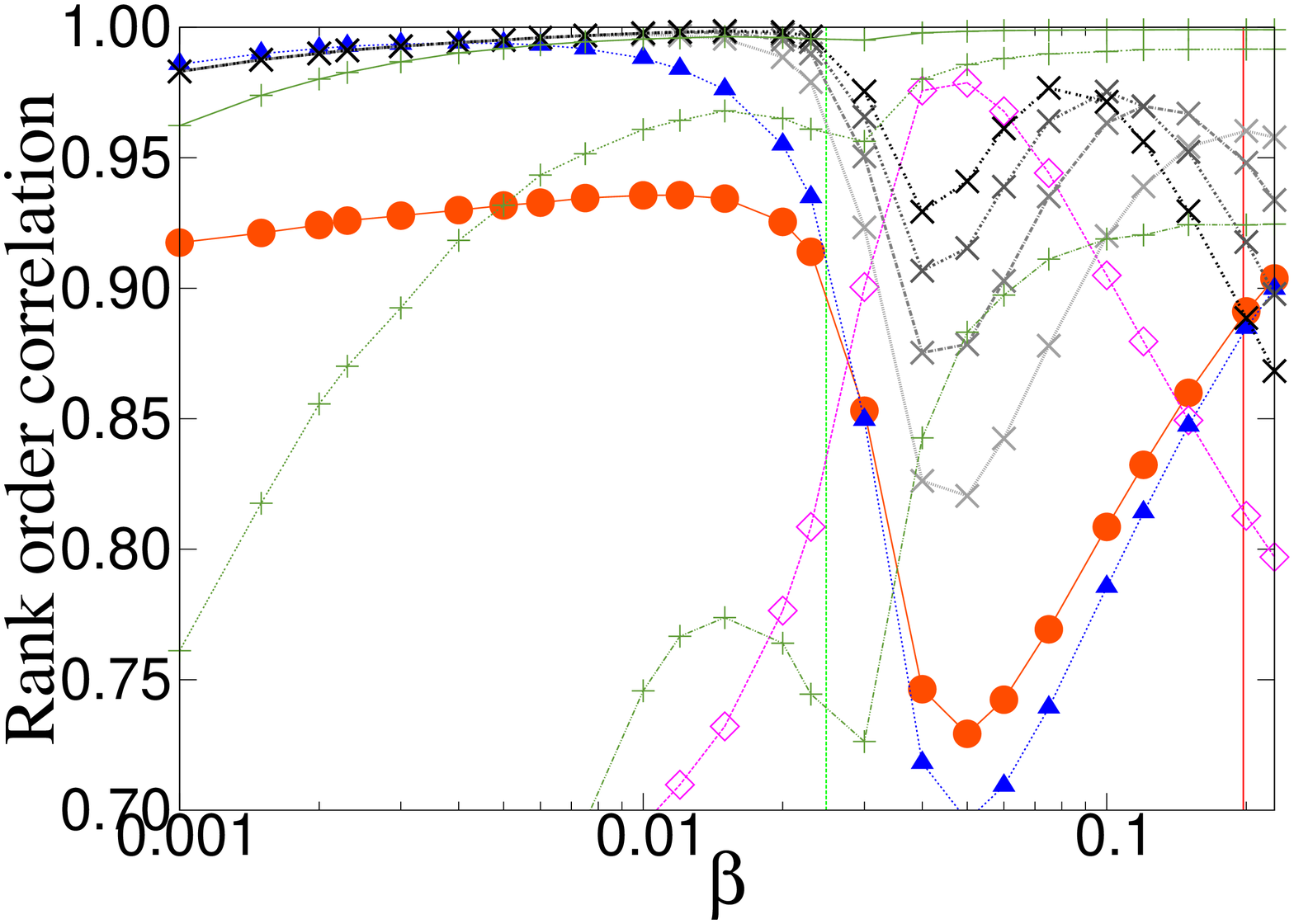}}
    \caption{\label{fig:correlationsFigs} Correlations and rank order correlations of various estimates of spreading effectiveness to simulated infection numbers on the structures of the email interaction, C. elegans neural and cond-mat networks. Results are plotted for estimates obtained using $k$-shells (red circles), eigenvector centrality (purple diamonds), out-degree or $I_i^{SIR}(L=1)$ (blue triangles), our estimation technique $I_i^{SIR}(L)$ using self-avoiding walks of length $L=2$ to 5 for cond-mat and 7 for the other networks (greyscale $\times$, darker gray-black to indicate longer walk lengths $L$), and for estimates from smaller numbers of simulations (10, 100 and 1000, which increase in accuracy) for the cond-mat network only (green $+$). Vertical (left) green lines indicate critical spreading at $\beta_c$ and (right) red lines indicate $\beta_{-3 dB}$ with 30 \% of network infected on average.}
\end{figure}

The correlation results for $I_i^{SIR}(L)$ generally improve as $L$ increases.
Estimates up to only short path lengths $L$ do not properly capture the effects of spreading on the network structure further away from patient zero when these nodes become more vulnerable at larger $\beta$.
In particular, $L=1$ captures only the out-degree of the initially infected node, and therefore does not represent any network structure more than one hop away.
In general then, one faces a trade-off between accuracy of inference of effectiveness against shorter computational time.
Importantly though, very good results can be obtained with short path lengths $L$, with the results from $L=4$ say being almost indistinguishable from longer $L$ for most of the range of $\beta$. This is a crucial point, since the runtime for the computations for $L \leq 4$ is much faster than simulations, and is on the order of the runtimes for the more simple degree ($L=1$) and $k$-shell inference methods for the small networks.
Finally, we note that the accuracy of the method drops once $\beta$ is well-inside the super-critical regime (even for large $L$) due to: i. insufficient path length at high $\beta$, ii. our independent walks assumption becomes less valid at high $L$ and $\beta$, and iii. with most nodes infecting a large proportion of the network, the structure surrounding each node makes less of an impact on the spreading efficiency.

Crucially, the accuracy achieved by our $I_i^{SIR}(L)$ can only be matched by large numbers of simulations, which take significantly longer runtime.
\fig{correlationsFigs} shows that, for the cond-mat network, using only 10 repeat simulations (with runtime double that of $I_i^{SIR}(L=5)$) provides much worse correlations, while comparable correlations up to the lower super-critical regime can only be achieved with 1000 simulations which cost around 200$\times$ more runtime.
As deduced earlier, our technique has asymptotically faster runtime by a significant constant factor.

Further, our estimates $I_i^{SIR}(L)$ are consistently more accurate with $L \geq 2$ than the $k$-shell inference, for $\beta$ values in the sub-critical, critical and early super-critical regimes.
A similar conclusion holds against the eigenvalue centrality measure of \cite{klemm11a}, for all but a couple of $\beta$ values near the critical regime in \fig{condmatRankOrder}; indeed, eigenvector centrality only achieves comparable accuracy near the critical regime.
These are crucial results covering the regimes of importance (since in the deeper super-critical regime, a large proportion of the network becomes infected and understanding the spreading efficiency of various initially infected nodes becomes less important).
Though our estimates are less efficient than $k$-shell or eigenvalue centrality, they are still much faster than simulations, and these results suggest a strong advantage to using $I_i^{SIR}(L)$.

\begin{figure*}[t]
        \subfigure[$\beta=0.023$]{\label{fig:actualVersusInferredBeta0.023}\includegraphics[width=0.24\textwidth]{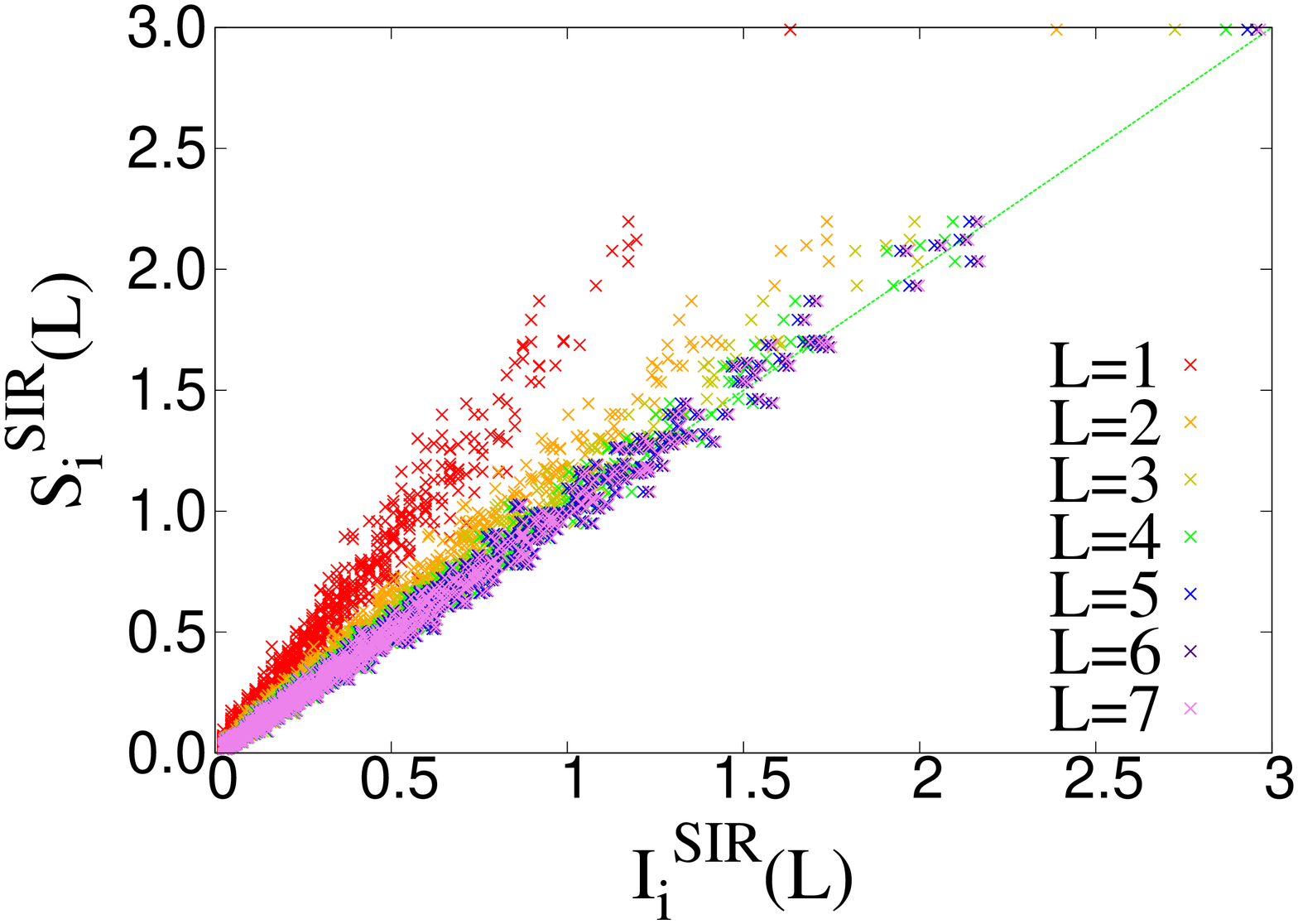}}
    \subfigure[$\beta=0.040$]{\label{fig:actualVersusInferredBeta0.040}\includegraphics[width=0.24\textwidth]{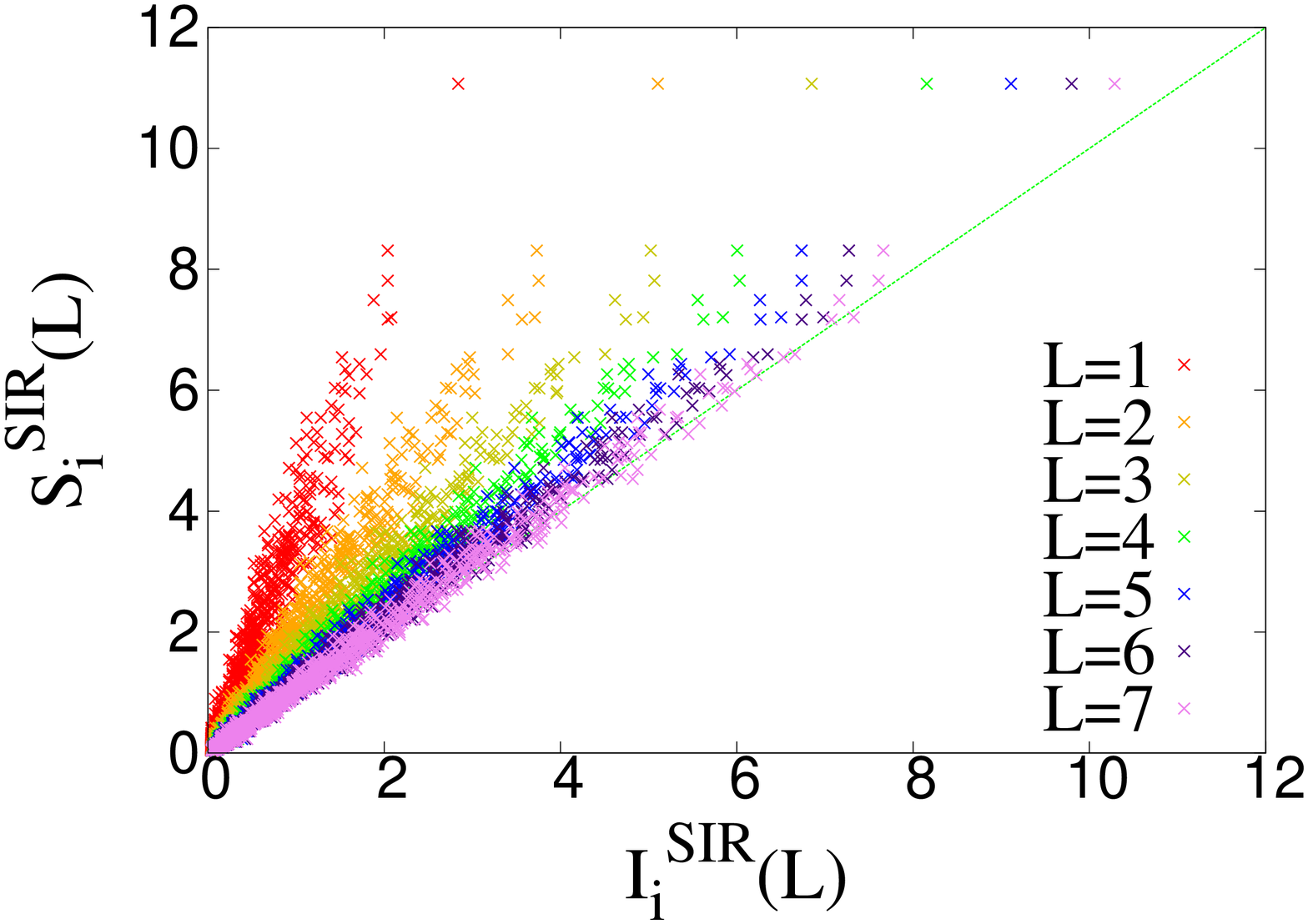}}
    \subfigure[$\beta=0.050$]{\label{fig:actualVersusInferredBeta0.050}\includegraphics[width=0.24\textwidth]{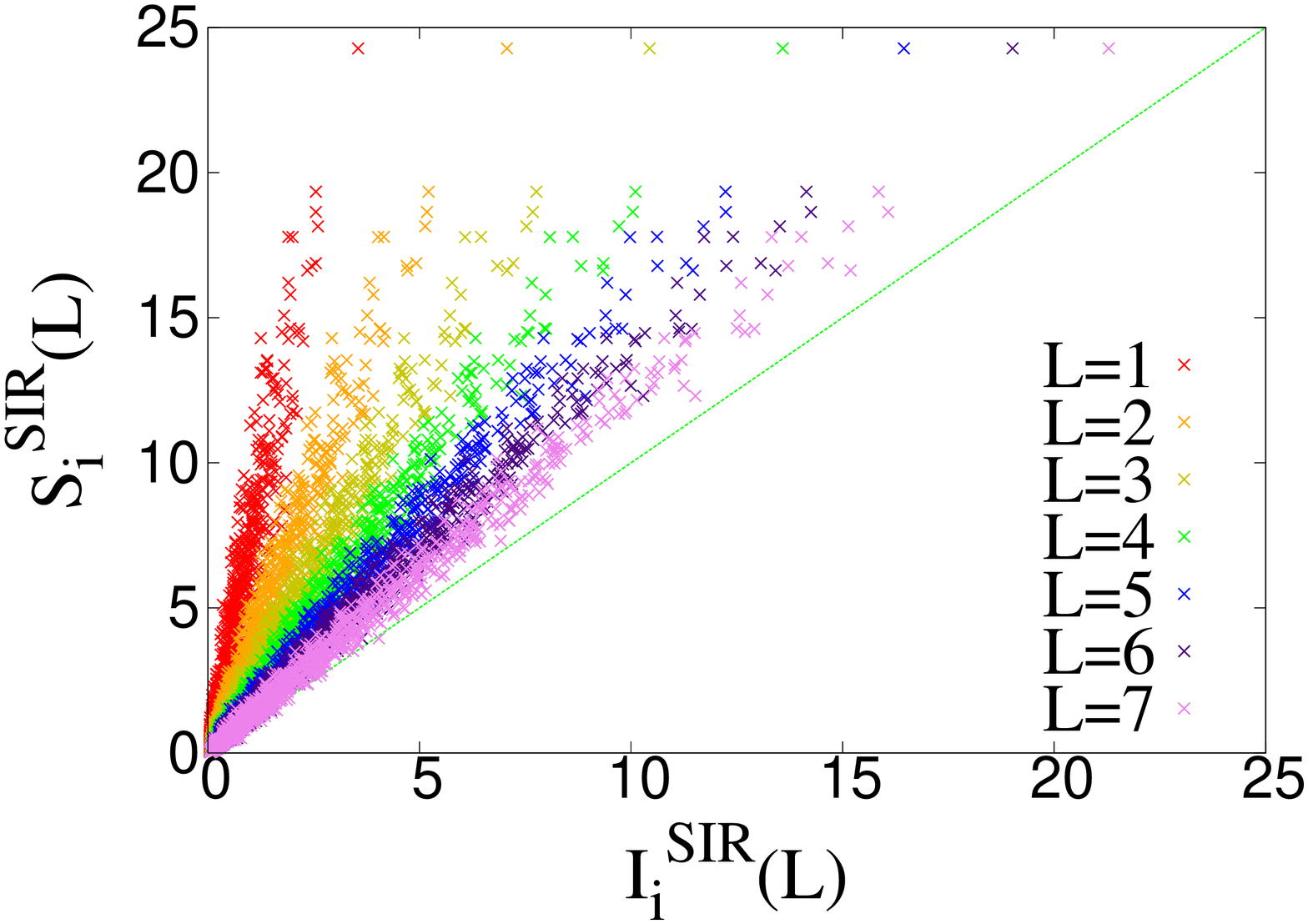}}
    \subfigure[$\beta=0.120$]{\label{fig:actualVersusInferredBeta0.120}\includegraphics[width=0.24\textwidth]{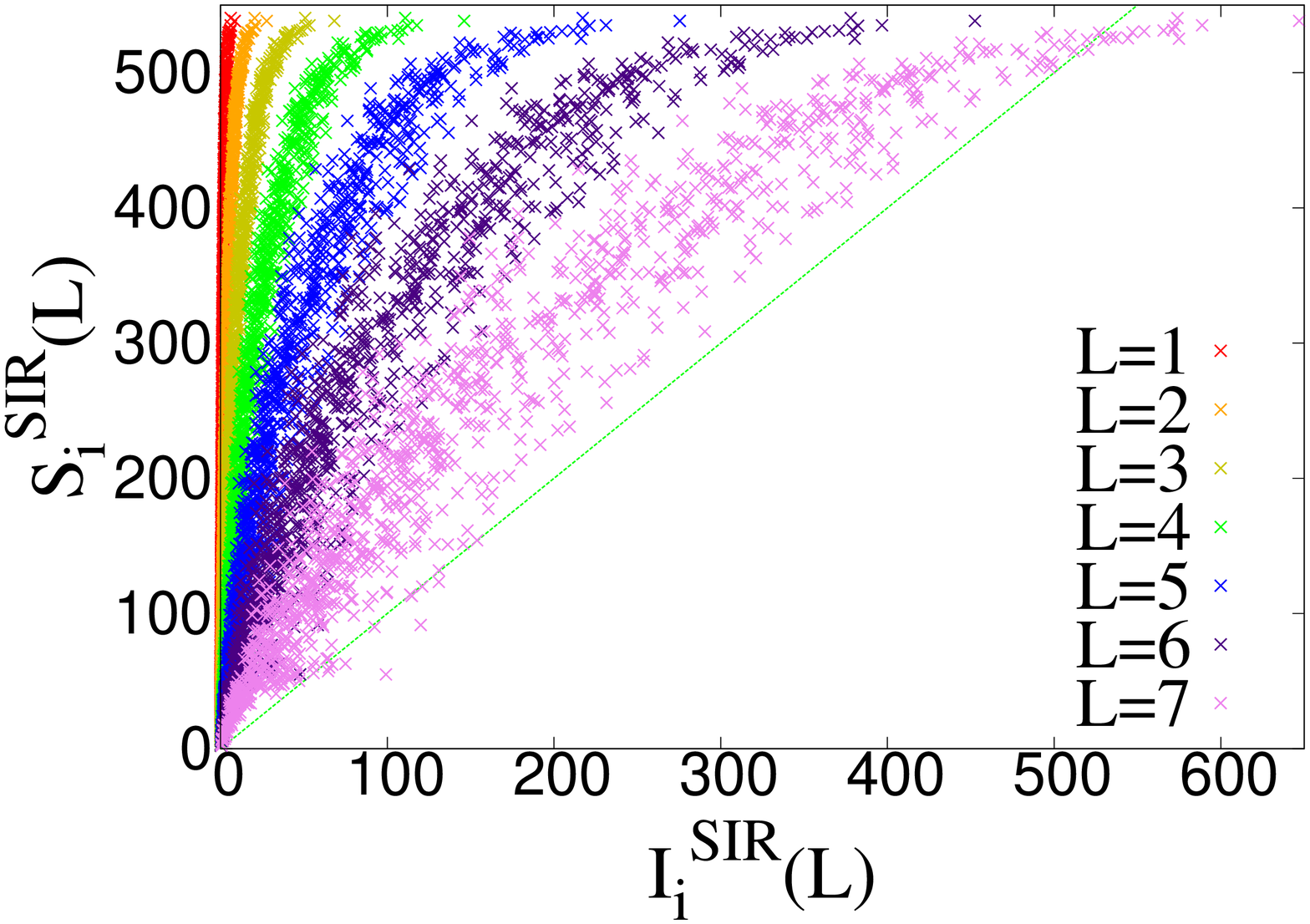}}
    \caption{\label{fig:actualVersusInferred} Simulated $S_i^{SIR}$ versus estimated $I_i^{SIR}(L)$ infection numbers, for each patient zero and maximum path lengths $L=1$ to 7, using the email interaction network structure.
    The straight green line represents the ideal plot $S_i^{SIR} = I_i^{SIR}(L)$.}
\end{figure*}

Additionally, we emphasise that while other tools such as the $k$-shell and eigenvector centrality are useful for inferring the \textit{relative} spreading effectiveness of each node, they do not actually estimate the number of infections from each node.
This is a distinct feature of our approach as compared to these other methods.
In \fig{actualVersusInferred} we directly compare our estimates $I_i^{SIR}(L)$ to the simulated values $S_i^{SIR}$ for each node $i$, for several values of $\beta$.
This clearly shows that our technique provides reasonable estimates of the expected number of infections across the sub-critical spreading regime and up to criticality for $L \geq 4$.
Indeed, reasonable accuracy can still be obtained with larger $L$ into the critical regime, though the time-efficiency benefits of doing so (as compared to simulation) declines.

\fig{actualVersusInferred} also demonstrates quite well the manner in which estimates improve (in general) with increasing maximum considered path length $L$.
We see that, while using too small a maximum path length $L$ appears to be the largest contributor to the inaccuracy of $I_i^{SIR}(L)$ (serving to pull points above the line $S_i^{SIR} = I_i^{SIR}(L)$), other errors are introduced by our approximations in \eq{plessq} and \eq{psusgreaterthanq} (the former of which pulls the points below this line by overestimating the probability of infection).
As previously stated though, for reasonable $\beta$ and large enough $L$, these errors seem to roughly cancel.
Importantly also, while the \textit{correlation} of estimates to simulated infection numbers may not always increase with $L$ in the supercritical regime, larger $L$ values provide consistently more accurate estimates of infection \textit{numbers} (e.g. see $\beta=0.12$ in \fig{correlationsFigs} and \fig{actualVersusInferred}).

Finally, we consider a simple heuristic to determine an appropriate $L$ length to use.
For potential infection walks from patient zero of length $L$, and for small $\beta$ in the sub-critical regime (in particular with $d_o \beta < 1$), one can make a naive estimate of the expected numbers of infections at length $L$ as $(d_o \beta)^L$, where $d_o$ is the average out-degree of the network.
One can then compute minimum $L$ values to keep $(d_o \beta)^L$ below a given value $r$.
For instance, in the email network (with $d_o=9.62$), using $r = 0.02$ suggests that with $L \geq 4$ and $\beta \leq 0.039$ we will only neglect infections on walk lengths where the expected number of infections was below $0.02$.
Of course, this is a simple estimate, neglecting the effects of dependent walks and making an implicit assumption that this $r=0.02$ is not large enough to significantly alter the number of infections; however it, along with observing from the diameters that many walks can be captured with $L=4$, helps to explain why $L=4$ provides good results even as $\beta$ approaches criticality.

\section{Conclusions}
%
We have presented a method for efficiently estimating the number of resulting infections from a given initially infected node in a network model.
This technique focusses on counting the number of functional walks to each candidate for infection, and in SIR models the only type of walks of interest are \textit{self-avoiding} walks.
Our technique is distinct from other recently proposed measures to infer spreading effectiveness of each node because it focusses specifically on disease spreading walks rather than general measures of network topology.
We demonstrated our technique to provide consistently more accurate inference of spreading effectiveness than other candidate techniques such as $k$-shells up to the lower super-critical regime.
This accuracy improvement is obtainable in reasonable computational time for SIR models, while still more accurate assessment can be obtained by increasing the computational time, and faster assessment can be made for SIS models.
Our technique is also distinguished in specifically estimating the number of infections, and we demonstrated that these estimates are reasonably accurate over a large range of $\beta$ for large enough values of the maximum counted walk lengths $L$.

\acknowledgments
The research leading to these results has received funding from the
European Research Council under the European Union's Seventh Framework
Programme (FP7/2007-2013) / ERC grant agreement n$^\circ$ 267087.
We thank the CSIRO High Performance Computing and Communications Centre 
and the Max Planck Institute for Mathematics in the Sciences for use of their computing clusters.

\bibliography{diseasespreading}

\begin{thebibliography}{10}
\expandafter\ifx\csname url\endcsname\relax\def\url#1{\texttt{#1}}\fi

\bibitem{Cohen01}
\Name{{Cohen R., Erez K., ben-Avraham D. and Havlin S.}} \REVIEW{Phys. Rev.
  Lett.}{86}{2001}{3682}.

\bibitem{Albert00}
\Name{Albert R., Jeong H. \and Barabasi A.} \REVIEW{Nature}{406}{2000}{378}.

\bibitem{Pastor-Satorras01}
\Name{Pastor-Satorras R. \and Vespignani A.} \REVIEW{Phys. Rev.
  Lett.}{86}{2001}{3200}.

\bibitem{Dodds04}
\Name{Dodds P.~S. \and Watts D.~J.} \REVIEW{Phys. Rev.
  Lett.}{92}{2004}{218701}.

\bibitem{Chen2011}
\Name{Chen D., L\"u L., Shang M.-S., Zhang Y.-C. \and Zhou T.} \REVIEW{Physica
  A}{391}{2012}{1777}.

\bibitem{Newman02}
\Name{Newman M. E.~J.} \REVIEW{Phys. Rev. E}{66}{2002}{016128}.

\bibitem{Chung09}
\Name{Chung F., Horn P. \and Lu L.} \Book{The giant component in a random
  subgraph of a given graph} Vol. 5427 of \emph{Lecture Notes in Computer
  Science} 2009 pp. 38--49.

\bibitem{Kitsak10}
\Name{Kitsak M., Gallos L., Havlin S., Lijeros F., Muchnik L., Stanley H. \and
  Makse H.} \REVIEW{Nature Physics}{6}{2010}{888}.

\bibitem{klemm11a}
\Name{Klemm K., Serrano M.~A., Eguiluz V.~M. \and San~Miguel M.}
  \REVIEW{Scientific Reports}{2}{2012}{292}.

\bibitem{Lu11}
\Name{L\"u L., Zhang Y.-C., Yeung C.~H. \and Zhou T.} \REVIEW{PLoS
  ONE}{6}{2011}{e21202}.

\bibitem{Zhou12}
\Name{Zhou Y.-B., L\"u L. \and Li M.} \REVIEW{New Journal of
  Physics}{14}{2012}{033033}.

\bibitem{Radicchi09}
\Name{Radicchi F., Fortunato S., Markines B. \and Vespignani A.} \REVIEW{Phys.
  Rev. E}{80}{2009}{056103}.

\bibitem{eub05}
\Name{Eubank S.} \REVIEW{Jap. J. of Infectious Diseases}{58}{2005}{S9}.

\bibitem{Madras93}
\Name{Madras N. \and Slade G.} \Book{The Self-Avoiding Walk} (Birkh\"auser)
  1993.

\bibitem{hayes98}
\Name{Hayes B.} \REVIEW{American Scientist}{86}{1998}{314+}.

\bibitem{gui03}
\Name{Guimer\`{a} R., Danon L., D\'{\i}az-Guilera A., Giralt F. \and Arenas A.}
  \REVIEW{Phys. Rev. E}{68}{2003}{065103+}.

\bibitem{white86}
\Name{White J., Southgate E., Thompson J. \and Brenner S.} \REVIEW{Phil. Trans.
  R. Soc. London}{314}{1986}{1}.

\bibitem{watts98}
\Name{Watts D.~J. \and Strogatz S.~H.} \REVIEW{Nature}{393}{1998}{440}.

\bibitem{new04}
\Name{Newman M. E.~J.} \REVIEW{Phys. Rev. E}{69}{2004}{066133}.

\end{thebibliography}
\bibliographystyle{eplbib}
\end{document}